\documentclass[reqno,nofootinbib,reprint]{revtex4-1}
\usepackage{amsmath}
\usepackage{amsthm}
\usepackage{amssymb,soul}
\usepackage{tikz}[2010/10/13]
\usepackage{breakurl}
\usepackage{ifpdf}
\input{Qcircuit}
\usepackage{mathtools}
\usepackage{hyperref}
\usepackage{amssymb,color}

\newcommand{\be}{\begin{equation}}
\newcommand{\ee}{\end{equation}}
\newcommand{\blue}[1]{{\color{blue}#1}}
\newcommand{\beq}{\begin{eqnarray}}
\newcommand{\eeq}{\end{eqnarray}}

\newcommand{\beqs}{\begin{eqnarray*}}
\newcommand{\eeqs}{\end{eqnarray*}}
\newcommand{\Max}{\ket{\rm{Max}}}
\newcommand{\GHZ}{\ket{\rm{GHZ}}}

\newcommand{\FS}{\mathfrak{F}_{s}}
\newcommand{\mn}{/3}
\newcommand{\nn}{/4}
\newcommand{\size}[1]{\fontsize{8pt}{\baselineskip}\selectfont{#1}}

\newcommand{\fmeasure}[5]{
\node at (#1-#3,#2-#4) {\size{$#5$}};
\draw (#1,#2) --(#1,#2-#4)  arc (0:180:#3) -- (#1-#3-#3,#2);
}

\newcommand{\fdoublemeasure}[6]
{
\fmeasure{#1}{#2}{3*#3}{#4}{}
\fmeasure{#1-2*#3}{#2}{#3}{#4}{#5}
\node at (#1-3*#3, #2-#4-2*#3) {\size{$#6$}};
}

\newcommand{\fqudit}[5]{
\node at (#1-#3,#2+#4) {\size{$#5$}};
\draw (#1,#2) --(#1,#2+#4)  arc (0:-180:#3) -- (#1-#3-#3,#2);
}

\newcommand{\fdoublequdit}[6]
{
\fqudit{#1}{#2}{3*#3}{#4}{}
\fqudit{#1-2*#3}{#2}{#3}{#4}{#5}
\node at (#1-3*#3, #2+#4+2*#3) {\size{$#6$}};
}

\newcommand{\fbraid}[4]{
\draw (#1,#2)--(#3,#4);
\draw (#1,#4)--(2/3*#1+1/3*#3,2/3*#4+1/3*#2);
\draw (#3,#2)--(2/3*#3+1/3*#1,2/3*#2+1/3*#4);
}

\newcommand{\fdoublebraid}[4]{
\draw (#1,#2)--(2/3*#3+1/3*#1,#4);
\draw (2/3*#1+1/3*#3,#2)--(#3,#4);
\draw (#1,#4)--(9/16*#1+7/16*2/3*#3+7/16*1/3*#1,9/16*#4+7/16*#2);
\draw (7/16*#1+9/16*2/3*#3+9/16*1/3*#1,7/16*#4+9/16*#2) --(5/16*#1+11/16*2/3*#3+11/16*1/3*#1,5/16*#4+11/16*#2);
\draw (3/16*#1+13/16*2/3*#3+13/16*1/3*#1,3/16*#4+13/16*#2) -- (2/3*#3+1/3*#1,#2);
\draw (#3,#2)--(9/16*#3+7/16*2/3*#1+7/16*1/3*#3,9/16*#2+7/16*#4);
\draw (7/16*#3+9/16*2/3*#1+9/16*1/3*#3,7/16*#2+9/16*#4) --(5/16*#3+11/16*2/3*#1+11/16*1/3*#3,5/16*#2+11/16*#4);
\draw (3/16*#3+13/16*2/3*#1+13/16*1/3*#3,3/16*#2+13/16*#4) -- (2/3*#1+1/3*#3,#4);
}

\begin{document}
\title{Compressed Teleportation}
\author{Arthur Jaffe}
\email{arthur\_jaffe@harvard.edu}
\author{Zhengwei Liu}
\email{zhengweiliu@fas.harvard.edu}
\author{Alex Wozniakowski}
\email{airwozz@gmail.com}
\affiliation{Harvard University, Cambridge, MA 02138, USA}
\begin{abstract}
In a previous paper we introduced holographic software for quantum networks, inspired by work on planar para algebras. This software suggests the definition of a compressed transformation.  Here we utilize the software to find a CT protocol to teleport compressed transformations. This protocol serves multiple parties with multiple persons.
\end{abstract}

\maketitle

\section{Introduction}
In a previous paper we introduced holographic software for quantum networks \cite{JLW}, inspired by work on planar para algebras \cite{JL}. We follow that notation: $X,Y,Z$ denote qudit Pauli matrices, $F$ is the Fourier transform, and $G$ is the Gaussian.  Recently, the quantum information community called for advances in teleportation, the ``most promising mechanism for a future quantum internet" \cite{PirandolaBraunstein}.  Here we utilize the holographic software to find a \textit{compressed teleportation} (CT) protocol.  This protocol serves multiple parties with multiple persons.

In our software we represent a $1$-qudit transformation as a ``two-string'' diagram, namely a diagram that has two input points and two output points. Many important transformations only act as  ``one-string'' transformations, such as Pauli matrices, or the controlled transformation on the control qudit. We call such transformations {\it compressed}.

An original teleportation protocol  was given by Bennett et al \cite{Bennett-etal}.  An optimized teleportation protocol was recently given in \cite{Yu-etal}, and in \cite{Yu} one finds extensive references.  All of these protocols are designed for two persons.

In this paper we give the new lossless CT protocol to teleport compressed transformations for multiple parties involving multiple persons.  This generalizes many teleportation protocols. Comparing CT with the  bidirectional teleportation for arbitrary transformations, our protocol reduces the cost of the resource state by 50\%. Even better, we only need one resource state for multiple persons, namely  $\Max$ introduced in \cite{JLW} as
\be
\Max=d^{\frac{1-n}{2}}\sum_{|\vec{k}|=0} \vec{\ket{k}}\;.
\ee
Here $\vec{\ket{k}}=(k_{1},\cdots,k_{n})$, with $k_{j}\in Z_{d}$, and $|\vec{k}|=\sum\limits_{j=1}^{n}k_{j}$.
On the other hand, the $\GHZ$ state, introduced in \cite{GHZ},
\be\label{GHZ}
\GHZ=d^{-\frac{1}{2}}\sum_{l=0}^{d-1}\ket{k,k,\cdots,k} \;,
\ee
is the Fourier transform of $\Max$. Namely
\be
\GHZ=(F\otimes \cdots \otimes F)\Max\;.
\ee
See \S I B and  \S III H  of \cite{JLW} for relations between the   Fourier transform $F$, string Fourier transform $\FS$, and entropy.

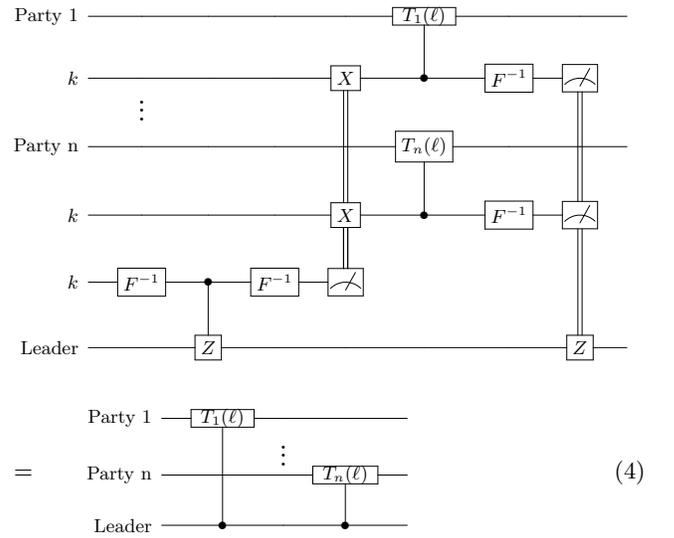
\begin{figure}[h]
\begin{align}
&\scalebox{0.8}{\raisebox{-2cm}{
\Qcircuit @C=1.5em @R=2em  {
\lstick{\text{Party 1}}&\qw&\qw&\qw& \qw&\multigate{0}{T_{1}(\ell)}&\qw&\qw&\qw \\
\lstick{k}& \dstick{ \Large{\vdots}} \qw & \qw & \qw & \gate{X} \cwx[2] & \control \qw \qwx[-1]  & \gate{F^{-1}}& \meter \cwx[2]\\
\lstick{\text{Party n}}&\qw&\qw&\qw&\qw&\gate{T_{n}(\ell)}&\qw&\qw&\qw\\
\lstick{k}& \qw & \qw & \qw & \gate{X} \cwx[1] & \control \qw \qwx[-1] & \gate{F^{-1}} & \meter \cwx[2]\\
\lstick{k}&\gate{F^{-1}} & \control \qw \qwx[1] & \gate{F^{-1}}    & \meter \cwx[-1]  \\
\lstick{\text{Leader}}& \qw & \gate{Z} & \qw   & \qw & \qw & \qw & \gate{Z} & \qw\\
}}}\nonumber\\
&\nonumber\\
=\quad\quad&\quad\quad\quad
\scalebox{0.8}{\raisebox{1cm}{
\Qcircuit @C=1.5em @R=2em  {
\lstick{\text{Party 1}}& \multigate{0}{T_{1}(\ell)}&\qw &\qw &\qw\\
\lstick{\text{Party n}}& \qw& \ustick{ \Large{\vdots}} \qw&\multigate{0}{T_{n}(\ell)}&\qw \\
\lstick{\text{Leader}}& \control \qw \qwx[-2] & \qw& \control \qw \qwx[-1] & \qw
}}}
\end{align}
\caption{CT protocol for controlled tranformations: The $k$'s arise from $\GHZ$ in \eqref{GHZ} for the leader and the persons $P_j$, $1\leq j\leq n$.
The output of the protocol is the multi-party-controlled transformation $T_c$.\label{Fig:Z teleportation}}
\end{figure}

\begin{figure}[h]
\scalebox{0.8}{
\Qcircuit @C=1.5em @R=2em  {
& \gate{T_j(\ell)} \qwx[1] \qw & \qw\\
& \control \qw & \qw
}}
\caption{Controlled transformations. \label{Fig:Controlled T}}
\end{figure}
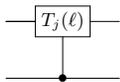

Let us describe our CT protocol in Fig.~\ref{Fig:Z teleportation}.
Suppose a network has one leader and $n$ parties. Also assume that the $j^{\rm th}$ party can perform a controlled transformation
\be
T_j=\sum_{l=0}^{d-1} \ket{\ell}\bra{\ell} \otimes T_{j}(\ell),
\ee
where the control qudit belongs to the person $P_j$, and $T_j(\ell)$ can be an arbitrary multi-person, multi-qudit transformation. (The protocol of the controlled transformation $T_j$ is shown in Fig.~\ref{Fig:Controlled T}.)
Under these conditions, they can perform a controlled transformation
\be
T_c=\sum_{l=0}^{d-1} \ket{\ell}\bra{\ell} \otimes T_{n}(\ell) \otimes \cdots \otimes T_{1}(\ell)\;,
\ee
for the network, using a resource state $\Max$ among the leader and the persons $P_j$.
The leader has the common control qudit in $T$, and the $j^{\rm th}$ party performs the transformation $T_{j,l}$ for control qudit $\ell$.
This protocol costs one resource state $\Max$ and $2n$ cdits. The time cost is the transmission of two cdits and the implementation of local transformations.

In \cite{JLW} we analyze the BVK protocol \cite{BVK} using holographic software. It would be interesting to analyze other protocols by this method, such as those  in \cite{bellexperiment,NielsenChuang,SorensenMolmer,GottesmanChuang,LoockBraunstein,Zhou-etal,Eisert-etal,Huelga-etal,Reznik-etal,ZhaoWang,LuoWang,Tele2015}.

\section{CT Details}
\begin{figure}[!htb]
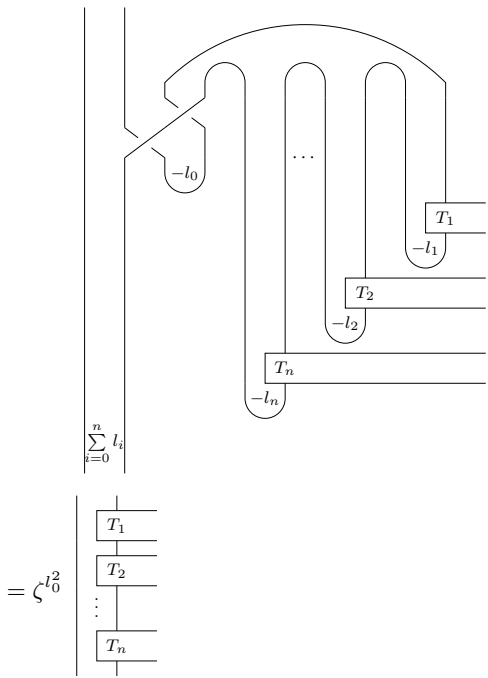

\begin{align*}
&~\scalebox{0.8}{\tikz{
\fbraid{6\mn}{0}{8\mn}{2\nn}
\fbraid{10\mn}{4\nn}{8\mn}{2\nn}
\draw (8\mn,4\nn) -- (8\mn,5\nn) to [bend left=45] (22\mn,5\nn)--(22\mn,4\nn);
\fqudit{18\mn}{4\nn}{-1\mn}{1\nn}{}{}
\fqudit{14\mn}{4\nn}{-1\mn}{1\nn}{}{}
\fqudit{10\mn}{4\nn}{-1\mn}{1\nn}{}{}
\fmeasure{8\mn}{0\nn}{-1\mn}{1\nn}{}
\node at (9\mn,-1\nn) {\size{$-l_0$}};
\draw (10\mn,0\nn) -- (10\mn,2\nn);
\node at (15\mn,0\nn) {$\cdots$};
\draw (22\mn,-5\nn+2\nn) -- (22\mn,4\nn);
\draw (20\mn,-7\nn+2\nn) -- (20\mn,4\nn);
\draw (24\mn,-5\nn+2\nn)--(21\mn,-5\nn+2\nn) -- (21\mn,-7\nn+2\nn) --(24\mn,-7\nn+2\nn);
\node at (22\mn,-6\nn+2\nn) {\size{$T_1$}};
\fmeasure{20\mn}{-7\nn+2\nn}{-1\mn}{1\nn}{-l_1}
\draw (-4\mn+22\mn,-5\nn-3\nn) -- (-4\mn+22\mn,4\nn);
\draw (-4\mn+20\mn,-7\nn-3\nn) -- (-4\mn+20\mn,4\nn);
\draw (24\mn,-5\nn-3\nn)--(-4\mn+21\mn,-5\nn-3\nn) -- (-4\mn+21\mn,-7\nn-3\nn) --(24\mn,-7\nn-3\nn);
\node at (-4\mn+22\mn,-6\nn-3\nn) {\size{$T_2$}};
\fmeasure{-4\mn+20\mn}{-7\nn-3\nn}{-1\mn}{1\nn}{-l_2}
\draw (-8\mn+22\mn,-5\nn-8\nn) -- (-8\mn+22\mn,4\nn);
\draw (-8\mn+20\mn,-7\nn-8\nn) -- (-8\mn+20\mn,4\nn);
\draw (24\mn,-5\nn-8\nn)--(-8\mn+21\mn,-5\nn-8\nn) -- (-8\mn+21\mn,-7\nn-8\nn) --(24\mn,-7\nn-8\nn);
\node at (-8\mn+22\mn,-6\nn-8\nn) {\size{$T_n$}};
\fmeasure{-8\mn+20\mn}{-7\nn-8\nn}{-1\mn}{1\nn}{-l_n}
\node at (5\mn,-19\nn) {\size{$\sum\limits_{i=0}^n l_i$}};
\draw (4\mn,10\nn) -- (4\mn,-21\nn);
\draw (6\mn,0) -- (6\mn,-21\nn);
\draw (6\mn,10\nn) -- (6\mn,2\nn);
}}\\
=\zeta^{l_0^2}\; &
\scalebox{0.8}{\raisebox{-1.3 cm}{
\tikz{
\draw (22\mn,-5\nn) -- (22\mn,-4\nn);
\draw (20\mn,-7\nn) -- (20\mn,-4\nn);
\draw (24\mn,-5\nn)--(21\mn,-5\nn) -- (21\mn,-7\nn) --(24\mn,-7\nn);
\node at (22\mn,-6\nn) {\size{$T_1$}};
\draw (22\mn,-5\nn-3\nn) -- (22\mn,-4\nn-3\nn);
\draw (20\mn,-7\nn-3\nn) -- (20\mn,-4\nn-3\nn);
\draw (24\mn,-5\nn-3\nn)--(21\mn,-5\nn-3\nn) -- (21\mn,-7\nn-3\nn) --(24\mn,-7\nn-3\nn);
\node at (22\mn,-6\nn-3\nn) {\size{$T_2$}};
\node at (21\mn,-11\nn) {\size{$\vdots$}};
\draw (22\mn,-5\nn-8\nn) -- (22\mn,-4\nn-6\nn);
\draw (20\mn,-7\nn-8\nn) -- (20\mn,-4\nn-6\nn);
\draw (22\mn,-5\nn-11\nn) -- (22\mn,-4\nn-11\nn);
\draw (20\mn,-5\nn-11\nn) -- (20\mn,-4\nn-11\nn);
\draw (24\mn,-5\nn-8\nn)--(21\mn,-5\nn-8\nn) -- (21\mn,-7\nn-8\nn) --(24\mn,-7\nn-8\nn);
\node at (22\mn,-6\nn-8\nn) {\size{$T_n$}};
}}}
\end{align*}
\caption{Diagrammatic CT-protocol for $X$-compressed transformations.} \label{CTX-Protocol-Pic}
\end{figure}
We say that a transformation $T$ is $Z$-compressed on the $i^{\rm th}$-qudit if $T$ is generated by Pauli $Z$ on the $i^{\rm th}$-qudit, and arbitrary transformations on the other qudits.  Similarly we define $X$-compressed or $Y$-compressed transformations.
Note that a transformation $T$ is $Z$-compressed on the first qudit if and only if it is a controlled transformation, namely $T=\sum\limits_{\ell=0}^{d-1} \ket{\ell}\bra{\ell}\otimes T(\ell)$.

We can switch between the three compressed transformations using
$FXF^{-1}=Z$ and $GXG^{-1}=Y^{-1}$; see \S II B of \cite{JLW} for details.

We say that a transformation $T'$ is compressed on the $i^{\rm th}$-qudit if $T'=UTV$, where $T$ is $Z$-compressed on the $i^{\rm th}$-qudit and  $U,V$ are local transformations on the $i^{\rm th}$-qudit.
We give the CT diagrammatic protocol for $X$-compressed transformations in Fig.~\ref{CTX-Protocol-Pic}.

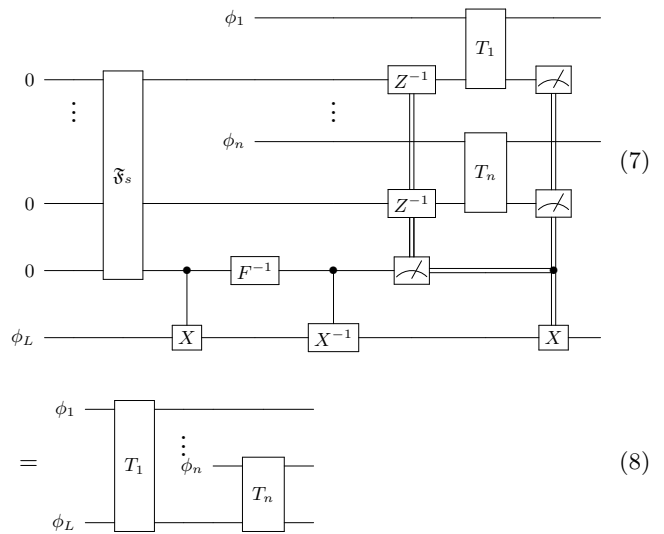
\begin{figure}[h]
\begin{align}
&\scalebox{0.8}{\raisebox{2.5cm}{
\Qcircuit @C=1.5em @R=2em  {
&&&&\lstick{\phi_1}&\qw&\qw& \multigate{1}{T_1}&\qw&\qw\\
\lstick{0}   & \dstick{ \Large{\vdots}} \qw                 & \multigate{3}{\FS} & \qw & \qw & \dstick{ \Large{\vdots}} \qw & \gate{Z^{-1}} \cwx[2] & \ghost{T_1}  & \meter \cwx[2]\\
&&&&\lstick{\phi_n}&\qw&\qw&\multigate{1}{T_n}&\qw&\qw\\
\lstick{0}  &  \qw &  \ghost{\FS}       & \qw & \qw & \qw & \gate{Z^{-1}} \cwx[1] & \ghost{T_n}  & \meter \cwx[1]\\
\lstick{0} & \qw                     & \ghost{\FS}  & \control \qw \qwx[1] & \gate{F^{-1}} & \control \qw \qwx[1]   & \meter \cwx[-1] & \cw & \control \cw \cwx[1] \\
\lstick{\phi_L}& \qw & \qw  & \gate{X} & \qw & \gate{X^{-1}} & \qw & \qw & \gate{X} & \qw\\
}}}\\
&\nonumber\\
=&\quad~~\scalebox{0.8}{\raisebox{1cm}{
\Qcircuit @C=1.5em @R=2em  {
\lstick{\phi_1}& \multigate{2}{T_1}&\qw&\qw &\qw &\qw\\
& & \ustick{ \Large{\vdots}} & \lstick{\phi_n} &\multigate{1}{T_n}&\qw \\
\lstick{\phi_L}& \ghost{T_1} & \qw  &\qw & \ghost{T_n} & \qw
}}}
\end{align}
\caption{CT protocol for $X$ compressed tranformations: The resource state $\Max$ is expressed as $\FS\vec{\ket{0}}$. One can simplify the protocol by Fig.~\ref{Trick}} \label{Fig:CTX-Protocol-Alg}
\end{figure}

Using our dictionary of the holographic software, we give the CT algebraic protocol in Fig.~\ref{Fig:CTX-Protocol-Alg}.

\begin{figure}[h]
\scalebox{0.8}{
\Qcircuit @C=1.5em @R=2em  {
& \control \qw \qwx[1]  & \meter \cwx[1] \\
& \gate{X^{-1}}         & \gate{X} & \qw\\
}}
=
\scalebox{0.8}{
\Qcircuit @C=1.5em @R=2em  {
&  \qw  & \meter  \\
& \qw   & \qw & \qw\\
}}
  \caption{\label{Trick}}
\end{figure}

Taking the conjugation of
local transformations, we obtain the CT protocol for compressed transformations. In particular, taking the conjugate of
the Fourier transform $F$, we obtain the CT protocol for $Z$-compressed transformations ( or controlled transformations) in Fig. \ref{Fig:Z teleportation}.

In the case with only two persons, the CT protocol says:
Assume that a quantum network can perform a transformation $T$, which is compressed on a 1-qudit belonging to a network member Alice. Then Alice can teleport her 1-qudit transformation to Bob using one edit and two cdits.
One can easily derive the swapping protocol, and the teleportation of the Tofolli gate from it.

\begin{acknowledgements}
\noindent This research was supported in part by a grant from the Templeton Religion Trust.  We are also grateful for hospitality at the FIM of the ETH-Zurich, where part of this work was carried out.
\end{acknowledgements}

\end{document}